\pdfoutput=1
\documentclass[aps,twocolumn, floatfix, prl,superscriptaddress]{revtex4-1}
\usepackage{graphicx}
\DeclareGraphicsExtensions{.pdf}
\usepackage{amsmath,amssymb,bbold,bm,color}
\usepackage{float}
\usepackage{epstopdf}

\newcommand{\br}{{\bm r}}

\newcommand{\bR}{{\bm R}}
\newcommand{\bB}{{\bm B}}

\newcommand{\bA}{{\bm A}}
\newcommand{\ba}{{\bm a}}
\newcommand{\bS}{{\bm S}}

\newcommand{\bdel}{{\bm \delta}}

\newcommand{\cH}{{\cal H}}

\newcommand{\bee}{\begin{equation}}
\newcommand{\ee}{\end{equation}}

\begin{document}

\title{Quantum holography in a graphene flake with an irregular boundary}

\author{Anffany Chen}
\affiliation{Department of Physics and Astronomy, University of
British Columbia, Vancouver, BC, Canada V6T 1Z1}
\affiliation{Quantum Matter Institute, University of British Columbia, Vancouver BC, Canada V6T 1Z4}
\author{R. Ilan}
\affiliation{Raymond and Beverly Sackler School of Physics and Astronomy,
Tel-Aviv University, Tel-Aviv 69978, Israel}
\author{F. de Juan}
\affiliation{Rudolf Peierls Centre for Theoretical Physics, Oxford, 1 Keble Road, OX1 3NP, United Kingdom}
\author{D.I. Pikulin}
\affiliation{Station Q, Microsoft Research, Santa Barbara, California 93106-6105, USA}
\author{M. Franz}
\affiliation{Department of Physics and Astronomy, University of
British Columbia, Vancouver, BC, Canada V6T 1Z1}
\affiliation{Quantum Matter Institute, University of British Columbia, Vancouver BC, Canada V6T 1Z4}

\begin{abstract} 
 Electrons in clean macroscopic samples of graphene exhibit an
astonishing variety of quantum phases when strong perpendicular
magnetic field is applied. These include integer and fractional
quantum Hall 
states
as well as symmetry broken phases and quantum Hall
ferromagnetism. 
Here we show that mesoscopic graphene flakes in the regime of strong
disorder and magnetic field can exhibit another remarkable quantum phase described
by holographic duality to an extremal black hole in two dimensional
anti-de Sitter space. This phase of matter can be characterized as a
maximally chaotic non-Fermi liquid 
since it is described by a complex fermion version of the 
Sachdev-Ye-Kitaev model known to possess these remarkable properties.
\end{abstract}

\date{\today}
\maketitle

Tensions between the laws of quantum mechanics and classical gravity
that are emblematic of the
extreme environments occurring in the early universe and near 
horizons of black holes constitute the most
enigmatic mysteries in modern physics. A promising avenue to
resolve some of the paradoxes encountered in these studies, such as
the black hole information paradox, is the holographic principle \cite{Bousso_RMP}. In
holographic duality,  quantum
gravity degrees of freedom in a $(d+1)$-dimensional spacetime ``bulk'' are
represented by a
many-body system defined on its $d$-dimensional boundary.

Important new insights into these fundamental questions have been gained
recently through the study of the  Sachdev-Ye-Kitaev (SYK) model
 \cite{SY1996,Kitaev2015} which
describes a system of $N$ fermions in (0+1) dimensions subject to random all-to-all
four-fermion interactions and is dual to dilaton gravity in (1+1) dimensional
anti-de Sitter space AdS$_2$  \cite{Sachdev2015,Maldacena2016}. Despite being maximally strongly interacting this
model is, remarkably, exactly solvable in the limit of large $N$. It
has been shown
to exhibit physical properties characteristic of the black hole,
including the extensive ground state entropy $S_0\sim N$, emergent
conformal symmetry at low energy and fast scrambling of quantum information that
saturates the fundamental bound on the relevant Lyapunov chaos
exponent $\lambda_T$.  Extensions of this model also
show interesting behaviors, including unusual spectral properties  \cite{Xu2016,Polchinski2016,Verbaar2016}, supersymmetry  \cite{Fu2016}, quantum phase transitions of an
unusual type  \cite{Altman2016, Bi2017,Lantagne2018}, quantum chaos propagation
 \cite{Gu2016,Berkooz2016,Hosur2016}, patterns of entanglement \cite{Liu2017,Huang2017} and strange metal behavior  \cite{Balents2017}.  
\begin{figure}[t]
\includegraphics[width = 8.6cm]{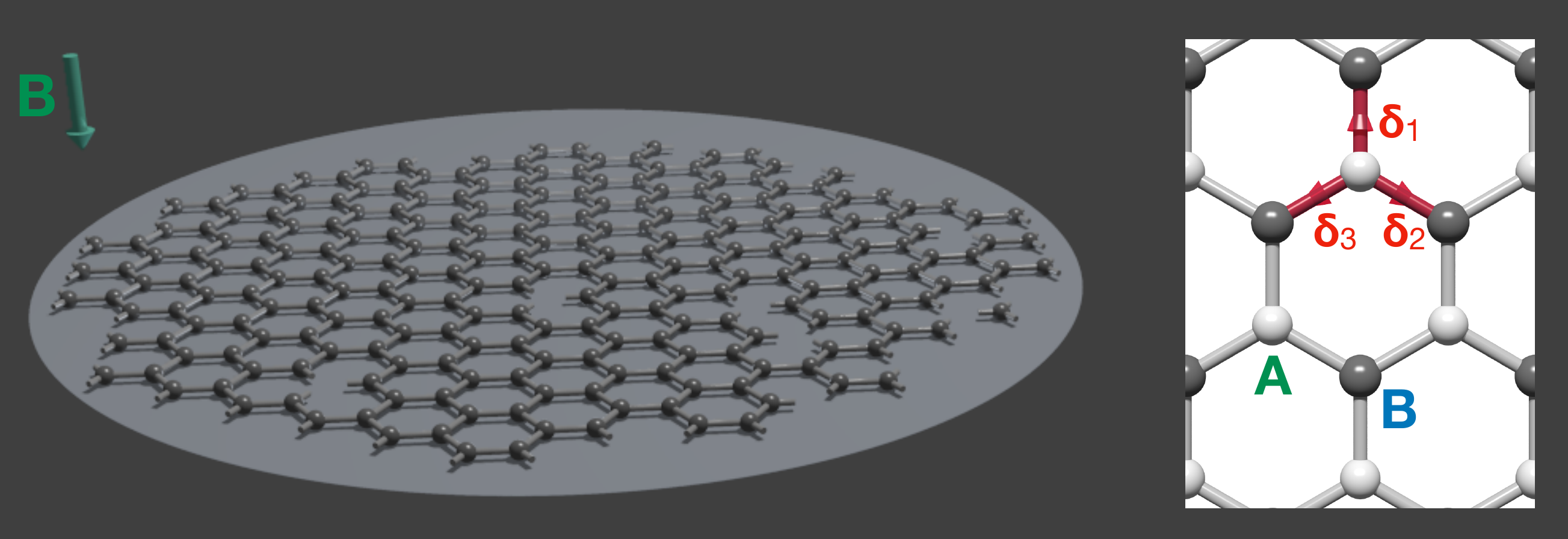}
\caption{{ Schematic depiction of the proposed device.}
  Irregular shaped graphene flake in applied magnetic field 
  $B$ forms the (0+1) dimensional many-body system equivalent to a
  black hole in (1+1) anti-de Sitter space. Inset: lattice structure of
  graphene with A and B sublattices marked and nearest neighbor
  vectors denoted by $\bdel_j$. 
}\label{fig1}
\end{figure}
\begin{figure*}[t]
\includegraphics[width = 17.6cm]{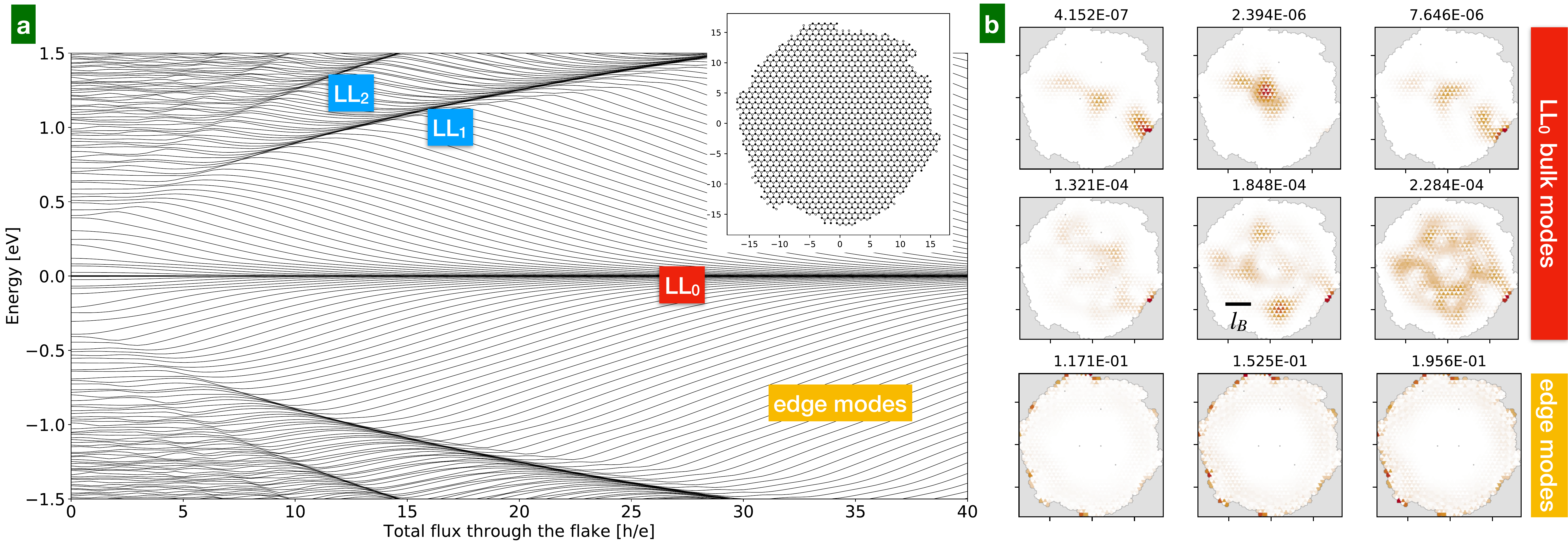}
\caption{{Electronic properties of an irregular graphene flake in
    the absence of interactions.}
 a) Single-particle energy levels $\epsilon_j$ of the Hamiltonian $H_0$ as a
 function of the magnetic flux $\Phi=SB$ through the flake. The flake  used for this calculation, 
depicted in the inset, consists of 1952 carbon atoms with equal number
of A and B sites. The
 energy spectrum, calculated here in the Landau gauge $\bA=Bx{\hat{\bm
     y}}$ and with open
 boundary conditions,  shows the same generic features irrespective of the
 detailed flake geometry. b) Typical wavefunction amplitudes of the
 eigenstates $\Phi_j(\br)$ belonging to LL$_0$ at $\Phi=40\Phi_0$ and
 the edge modes. The numerals above each panel
 denote the energy $\epsilon_j$  of the state in eV, scale bar shows the
 magnetic length $l_B=\sqrt{\hbar c/eB}$.
}\label{fig2}
\end{figure*}
In this letter we propose a simple experimental realization of the SYK model with
complex fermions in a mesoscopic graphene flake with an irregular
boundary and subject to a strong applied magnetic field. Unlike the earlier proposals in
solid state systems  \cite{Pikulin2017,Alicea2017}, which targeted the Majorana fermion version of
the model, our proposed device does not require superconductivity or
advanced fabrication techniques and should therefore be relatively
straightforward to assemble using only the existing technologies. The
proposed design is
illustrated in Fig.\ \ref{fig1}.  Magnetic field $B$ applied to
graphene is known to produce a variety of interesting quantum phases 
 \cite{Novoselov2005,Zhang2005,Novoselov1379,Bolotin2009,Du2009,Dean2011,Feldman1196,Nomura2006,Alicea2006,Amet2015}.
At the noninteracting level the field simply  reorganizes the single-particle
electron states into Dirac Landau levels with energies  \cite{Neto_RMP} $E_n\simeq
\pm\hbar v\sqrt{2n(eB/\hbar c)}$ and $n=0,1,\cdots$. We argue that
when the graphene flake is sufficiently small and irregular 
the electrons in the  $n=0$
Landau level (LL$_0$) are generically described by the SYK model. 
This remarkable property is rooted in the
celebrated Aharonov-Casher construction  \cite{Casher1979} which implies that, in the
absence of interactions, LL$_0$ remains
perfectly sharp even in the presence of strong disorder that respects the chiral symmetry of graphene.
As we shall see a flake with a highly irregular boundary, illustrated in
Fig.\ \ref{fig1}, is chirally symmetric.  Electrons in LL$_0$, therefore,
remain nearly perfectly degenerate, despite the fact that their
wavefunctions acquire random spatial structure. When Coulomb
repulsion is projected onto these highly disordered states,  
random all-to-all interactions between the zero modes are generated,
exactly as required to define the SYK model.

The  complex fermion SYK model, also known as the Sachdev-Ye (SY)
model  \cite{SY1996,French70,Bohigas71a,Bohigas71b},  is defined  by the second-quantized Hamiltonian
\begin{equation}\label{h1}
\cH_{\rm SY}=\sum_{ij;kl} J_{ij;kl}c^\dagger_ic^\dagger_jc_kc_l -\mu\sum_jc^\dagger_jc_j,
\end{equation}
where $c^\dagger_j$ creates a spinless fermion, $ J_{ij;kl}$ are
zero-mean complex random variables satisfying $ J_{ij;kl}=J^*_{kl;ij}$ 
and $J_{ij;kl}=-J_{ji;kl}=-J_{ij;lk}$ and $\mu$ denotes the chemical potential.
In what follows we derive the effective low-energy
Hamiltonian for electrons in LL$_0$ of a graphene flake with an
irregular boundary  and show that, under a broad range of conditions, it is given by Eq.\
(\ref{h1}). The system, therefore,
realizes the SY model.

At the non-interacting level a flake of graphene is described by a
simple tight-binding Hamiltonian \cite{Neto_RMP}
\begin{equation}\label{h2}
H_0=-t\sum_{\br,\bdel} (a^\dagger_\br b_{\br+\bdel}+{\rm h.c.}),
\end{equation}
where $a^\dagger_\br\ (b^\dagger_{\br+\bdel})$ denotes the creation
operator of the electron on the subblatice  A (B)  of the
honeycomb lattice. These satisfy the canonical anticommutation
relations $\{a^\dagger_\br,a_{\br'}\}=\{b^\dagger_\br,b_{\br'}\}=\delta_{\br\br'}$ appropriate
for fermion operators. $\br$ extends
over the sites in sublattice A while $\bdel$ denotes the 3 nearest
neighbor vectors (inset Fig.\ \ref{fig1}).   $t=2.7$ eV is the
tunneling amplitude \cite{Song2010}.  For simplicity we first
ignore electron spin but reintroduce it later. 
The chiral symmetry $\chi$ is generated by setting
$(a_\br,b_\br){\rightarrow} (-a_\br,b_\br)$ for all $\br$ which has the effect of flipping the
sign of the Hamiltonian $H_0\to -H_0$. 

Magnetic field $B$ is incorporated in the Hamiltonian (\ref{h2}) by means
of the standard  Peierls substitution which replaces $t\to t_{\br,\br+\bdel}=t
\exp{[-i(e/\hbar c)\int_\br^{\br+\bdel}\bA\cdot{d \bm l}]}$ where
$\bA$ is the vector potential $\bB=\nabla\times\bA$.
 In the presence of $\chi$ the
Aharonov-Casher construction  \cite{Casher1979} implies $N=N_\Phi$
exact zero modes in the spectrum of $H_0$ where
 $N_\Phi=SB/\Phi_0$ denotes the number of magnetic flux
quanta $\Phi_0=hc/e$ piercing the area $S$  of the flake. It is
  clear that a flake with an arbitrary shape described by $H_0$ respects
  $\chi$ which underlies the robustness of LL$_0$ invoked above. 

Hopping $t'$ between second neighbor sites and random on-site potential are
examples of perturbations that break $\chi$ and are therefore expected
to broaden LL$_0$. These
effects can be modeled  by adding to $\cH_{\rm SY}$ defined in Eq.\
(\ref{h1}) a term 
\begin{equation}\label{h22}
\cH_2=\sum_{ij}K_{ij}c^\dagger_ic_j
\end{equation}
 which
describes  a small (undesirable)  hybridization between the states in
LL$_0$ that will generically be present in any realistic experimental
realization. We discuss the effect of these terms below.

In Fig.\ \ref{fig2}a we show the single-particle energy spectrum of $H_0$ for a graphene flake
with a  shape depicted in the inset. As a function of
increasing magnetic field $B$ we observe new levels joining the
zero-energy manifold  LL$_0$ such that the number of zero modes
follows $N\simeq N_\Phi$ in accordance with the 
Aharonov-Casher argument. Higher Landau levels and topologically
protected edge modes are also visible. Despite the randomness
introduced by the irregular boundary LL$_0$ remains sharp as expected on the basis
of the arguments presented above. This is the key feature in our
construction of the SY Hamiltonian which guarantees that the $\cH_2$
term defined above vanishes as long as the chiral symmetry is respected. In
the presence of e-e repulsion the leading term in the effective
description of LL$_0$ will therefore be a four-fermion interaction which we discuss next.

\begin{figure}[t]
\includegraphics[width = 8.6cm]{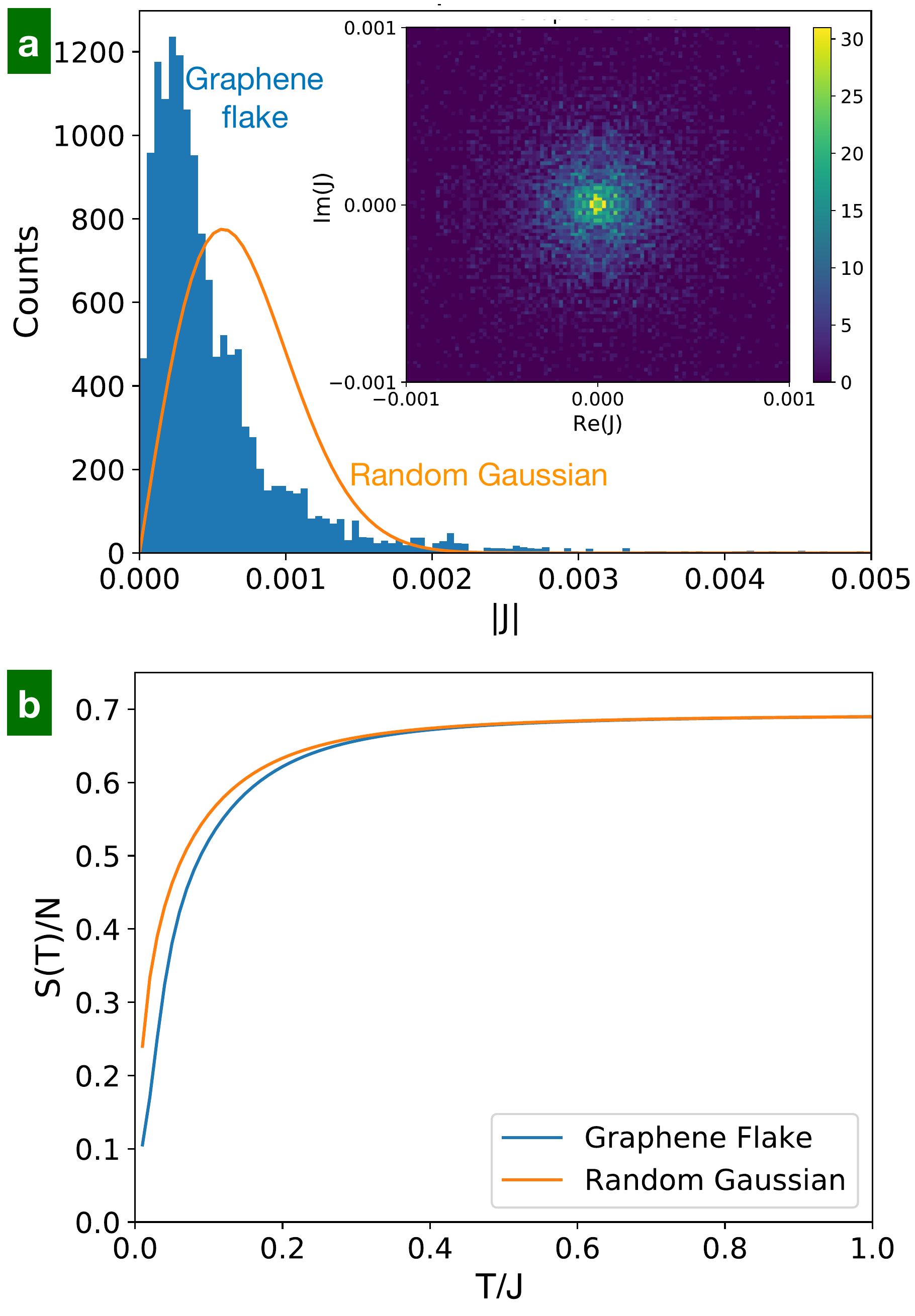}
\caption{{Statistical properties of the coupling constants and the
    thermal entropy.} 
a) Histogram of $|J_{ij;kl}|$ as calculated from Eq.\ (\ref{J1}) with $V_1=1$ for the
graphene flake depicted in Fig.\ \ref{fig2} and $N=16$, compared to the
Gaussian distribution (orange line) with the same variance $0.000805 V_1$. Inset shows
the histogram of real and imaginary components of
$J_{ij;kl}$. The mirror symmetry about the horizontal follows from the
hermiticity property  $ J_{ij;kl}=J^*_{kl;ij}$. b) Entropy $S(T)$ of
the SY Hamiltonian (\ref{h1}) calculated with $J$s shown in panel (a).
}\label{fig3}
\end{figure}
Electron wavefunctions $\Phi_j(\br)$ belonging to LL$_0$  exhibit
random spatial structure (Fig.\ \ref{fig2}b) owing to the irregular confining
geometry imposed by the shape of the flake. 
From the knowledge of the
wavefunctions it is straightforward to evaluate the corresponding interaction matrix
elements (Supplementary Section A) \footnote{See Supplemental Material
  [url] for details of this and onther calculations} between the zero modes.
 The leading many-body Hamiltonian for electrons in LL$_0$
will thus have the form of Eq.\ (\ref{h1}) with 
\begin{equation}\label{J1}
J_{ij;kl}={1\over 2}\sum_{\br_1,\br_2}[\Phi_i(\br_1)\Phi_j(\br_2)]^*V(\br_1-\br_2)[\Phi_k(\br_1) \Phi_l(\br_2)],
\end{equation}
where $V(\br)=(e^2/\epsilon r)e^{-r/\lambda_{TF}}$ is the
screened Coulomb potential with Thomas-Fermi length
$\lambda_{TF}$ and dielectric constant $\epsilon$. The summation extends
over all sites of the honeycomb lattice. It is to be noted that only
the part of $J_{ij;kl}$ antisymmetric in $(i,j)$ and $(k,l)$
contributes to the many-body Hamiltonian (\ref{h1}) so in the
following we assume that $J_{ij;kl}$ has been properly
antisymmetrized. 

 We numerically evaluated $J_{ij;kl}$ for various values of
 $\lambda_{TF}$. The resulting
 $J$s are complex valued random variables satisfying
\begin{equation}\label{J2}
\overline{J_{ij;kl}}=0, \ \ \  \overline{|J_{ij;kl}|^2}={1\over 2N^3} J^2,
\end{equation}
where $J$ measures the interaction strength and the bar denotes
averaging over randomness introduced by the irregular confining
geometry.  Fig.\ \ref{fig3}a shows the statistical distribution of 
$J_{ij;kl}$ calculated for the nearest-neighbor
 interactions $V(\br)=V_1\sum_{\bdel}\delta_{\br,\bdel}$ and the
 single-particle wavefunctions $\Phi_j(\br)$ depicted in Fig.\ \ref{fig2}b.
The distribution of $J_{ij;kl}$  shows the expected randomness  with
some deviations from the ideal Gaussian. 

To ascertain the effect of these deviations and to prove that the low-energy fermions in the graphene
flake are described by the SY model we  next perform numerical
diagonalization of the many-body Hamiltonian  (\ref{h1}) with coupling
constants  $J_{ij;kl}$ obtained as described above. 
We then calculate
various physical observables and compare them to the results obtained
with random independent  $J_{ij;kl}$. Fig.\ \ref{fig3}b shows the thermal entropy
$S(T)$ of the flake. Comparison to the entropy calculated with random
Gaussian $J_{ij;kl}$ indicates no significant difference. It is to be
noted that while the SY model is known to exhibit non-zero ground
state entropy per particle in the thermodynamic limit, $S(T)$ still
vanishes as $T\to 0$ for any finite $N$  \cite{Wengbo2016}. 

%
\begin{table}\label{ttable1}
\begin{tabular*}{0.48\textwidth}{@{\extracolsep{\fill}}l | c c c c }
\hline \hline 
 $N(\!\!\mod{4})$  & 0  & 1 & 2 & 3  \\
\hline
$q=0$ & GOE &  & GSE &  \\
$q\neq 0$ & GUE & GUE & GUE & GUE \\
\vspace{-6pt}\\
\hline \hline
\end{tabular*}
\caption{{\bf Gaussian ensembles for the SY model.} The relevant probability
  distributions are given by Eq.\ (\ref{syk4}) with $Z={8\over  27},{4\pi\over 81 \sqrt{3}},{4\pi\over 729\sqrt{3}} $ and
$\beta=1,2,4$ for GOE, GUE, GSE, respectively.}
\end{table}
Many-body energy level statistics provide another useful tool to
validate our hypothesis that LL$_0$ electrons in the graphene flake
behave according to the SY model. We thus arrange the
energy eigenvalues $E_n$ of the many-body Hamiltonian (\ref{h1}) 
in increasing order and form
ratios of the subsequent levels $r_n=(E_{n+1}-E_n)/(E_n-E_{n-1})$. 
According to the random matrix theory applied to the SY model  \cite{Xu2016} 
probability distributions  $P(\{r_n\})$ are given by different Gaussian
ensembles, depending on $N(\!\!\mod 4)$ and the eigenvalue $q$ of the
total charge operator $Q=\sum_j(c^\dagger_j c_j-1/2)$ as summarized in
Table I. 
Here GOE, GUE and GSE stand for Gaussian orthogonal, unitary and
symplectic ensembles, respectively and 
\begin{equation}\label{syk4}
P(r)={1\over Z}{(r+r^2)^\beta\over (1+r+r^2)^{1+3\beta/2}},
\end{equation}
with constants $Z$ and $\beta$ listed in Table I. Since $\cH_{\rm SY}$
commutes with $Q$ it can be block
diagonalized in sectors
with definite charge eigenvalue $q$. As emphasized in Ref.\ \onlinecite{Xu2016}
the level statistics analysis must be performed separately
for each $q$-sector. Note that $q$ has integer (half-integer) values for
$N$ even (odd) and this is why the neutrality condition $q=0$  can be met
only for even values of $N$. Also note that $q=0$ corresponds to $N/2$ particles.
\begin{figure}[t]
\includegraphics[width = 8.6cm]{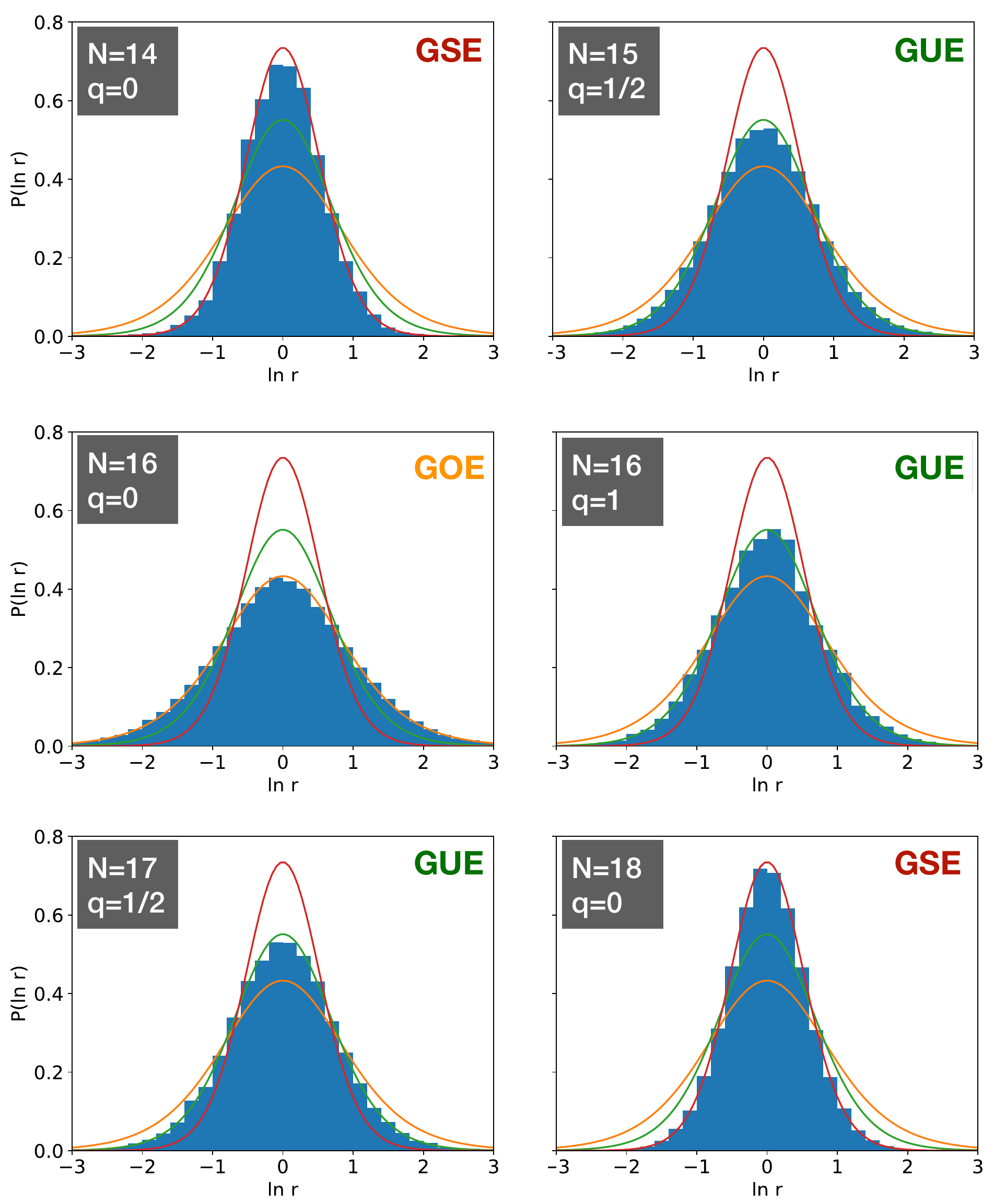}
\caption{{ Many-body level statistics for the interacting electrons
    in LL$_0$ of the graphene flake.} Blue bars show the calculated
  distributions for the graphene flake. Orange, green and red curves
  indicate the expected distributions given by Eq.\ (\ref{syk4}) for
  GOE, GUE and GSE, respectively. To obtain smooth distributions, 
  results for  
  $N=14,15,(16)$  have been averaged over 8 (4) distinct flake geometry 
  realizations while $N=17,18$ reflect a single realization.  
}\label{fig4}
\end{figure}

Fig.\ \ref{fig4} shows our results for the level statistics performed for a
graphene flake with 
$N=14$ through $18$ and various values of $q$. The obtained level
spacing distributions
are seen to unambiguously follow the prediction of the random matrix
theory for the SY model summarized in Table I.  We are thus led to conclude that
interacting electrons in LL$_0$ of a graphene flake with an irregular boundary
indeed exhibit spectral properties characteristic of the SY model. 

In the rest of this Letter we discuss various aspects of the problem
relevant to the laboratory realization.  Electrons in graphene possess
spin which we so far ignored. Given the weak spin-orbit coupling in
graphene we may model the non-interacting system by two copies of
the Hamiltonian Eq.\ (\ref{h2}) plus the Zeeman term,
$H=H_0+g^*\mu_B\bB\cdot\bS_{\rm tot}$ where $\bS_{\rm tot}$ is the
total spin operator and
$\mu_B=5.78\times 10^{-5}$ eV/T is the Bohr magneton. For graphene on
the  SiO$_2$ or hBN substrate we may take  $g^*\simeq 2$ which gives the bare Zeeman
splitting $\Delta E_S(B)\simeq 0.12$ meV/T, or about 2.4 meV at $B=20$
T.  We expect this relatively small spin splitting to be significantly
enhanced by the exchange effect of the  Coulomb repulsion. The
strength of the exchange splitting $\Delta E_C\simeq 8.8$ meV/T  is
 estimated in the Supplementary
Section A.  For such a large spin splitting one may focus on a
partially filled LL$_0$ for
a single spin projection and disregard the other. The spinless
model considered so far should therefore serve as an excellent approximation
of the physical system in the strong field.

Disorder that breaks chiral symmetry will inevitably be present in
real graphene samples. Such disorder tends to broaden LL$_0$ and compete
with the interaction effects that underlie the SY physics. It is
known that bilinear terms $\cH_2$ that arise from such disorder
constitute a relevant perturbation to  $\cH_{\rm SY}$ and drive the
system towards a disordered Fermi liquid (dFL) ground state. In the
Supplementary Section B we
analyze the symmetry-breaking effects and estimate their
strength in realistic situations.
We conclude that in carefully prepared samples a significant window
should remain open at non-zero temperatures and frequencies in which the
system exhibits behavior characteristic of the SY model.

An ideal sample to observe the SY physics is a graphene flake with a
highly irregular boundary and clean interior. These conditions
promote random spatial structure of the electron wavefunctions and
preserve degeneracy of LL$_0$. Disordered
wavefunctions give rise to random interaction matrix elements
$J_{ij;kl}$ while near-degeneracy of states in LL$_0$ guarantees that the
two-fermion term $\cH_2$ remains small. To observe signatures
of the emergent black hole the LL$_0$ degeneracy $N=SB/\Phi_0$ must be
reasonably large -- numerical simulations indicate that $N\gtrsim 10$
is required for the system to start
showing the characteristic spectral features. Aiming at $N\simeq 100$,
which is well beyond what one can conceivably simulate on a computer,
implies the characteristic sample size
$L\simeq\sqrt{S}=\sqrt{N\Phi_0/B}\simeq 150$ nm at $B=20$
T. Signatures of the SY physics can be observed spectroscopically,
e.g.\ by the differential tunneling
conductance $g(V)=dI/dV$ which is predicted \cite{Pikulin2017} to exhibit a
characteristic square-root divergence $g(V)\sim |V|^{-1/2}$ in the SY
regime at large $N$, easily distinguishable from the dFL behavior
$g(V)\sim$ const at  small $V$. We predict that a tunneling
experiment will observe the SY behavior when the chemical potential
of the flake is tuned to lie in LL$_0$ and dFL behavior for all LL$_n$ with
$n\neq 0$. We also expect the two-terminal conductance across the
flake to show interesting behavior in the SY regime but we defer a
detailed discussion of this to future work.

In the limit of a large flake the irregular boundary will eventually 
become unimportant for the electrons in the bulk interior and the system 
should undergo a crossover to a more conventional `clean'
phenomenology characteristic of graphene in applied magnetic
field. The exact nature of this crossover poses an interesting
theoretical as well as experimental problem which we also leave to
future study.  

{\em Acknowledgments --} The authors are indebted to Ian Affleck, 
Oguzhan Can, \'Etienne
Lantagne-Hurtubise and  Chengshu Li for stimulating discussions. The work
was supported by NSERC, CIfAR (AC,MF) and by the Marie Curie Programme
under EC Grant agreement No. 705968 (FJ)”.  Tight-binding simulations
were performed using Kwant code~ \cite{groth2014kwant} and
computational resources provided by WestGrid.

\bibliography{SYK}


\newpage

\renewcommand{\thepage}{S\arabic{page}} 
\renewcommand{\thetable}{S\arabic{table}}  
\renewcommand{\thefigure}{S\arabic{figure}} 
\renewcommand{\theequation}{S\arabic{equation}} 
\setcounter{page}{1}
\setcounter{equation}{0}
\setcounter{figure}{0}

\section{Supplementary material}
\subsection{Exchange splitting and the interaction matrix elements}

In this section we discuss the enhancement of the Zeeman splitting due
to the exchange interaction, derive the form of coupling constants
$J_{ij;kl}$ quoted in Eq.\ (4) of the main text and estimate the
characteristic interaction strength $J$. 

\subsubsection{General considerations}
We begin by writing the
Hamiltonian for the electrons in graphene as $H=H_0+H_{\rm int}$ where 
\begin{equation}\label{int1}
H_0=-\sum_{\langle\br,\br'\rangle,\sigma}t_{\br\br'}f_{\br\sigma}^\dagger f_ {\br'\sigma} +{g^*\mu_B
       B\over 2}\sum_{\br}(\rho_{\br\uparrow}-\rho_{\br\downarrow}), 
\end{equation}
Here $f_{\br\sigma}^\dagger$ creates an electron with spin $\sigma$ on
the site $\br$ of the honeycomb lattice and satisfies
$\{f_{\br\sigma}^\dagger,f_{\br'\sigma'}\}=\delta_{\br\br'}\delta_{\sigma\sigma'}$. Relative
to our notation in Eq.\ (2) of the main text we added the spin degree
of freedom. Aside from the spin degree of freedom $f_{\br\sigma}$ coincides with
$a_\br$ ($b_\br$) when $\br$ is in subblattice A (B). Peierls
substitution implies  $t_{\br\br'}=t
\exp{[-i(e/\hbar c)\int_\br^{\br'}\bA\cdot{d \bm l}]}$ for the hopping integral in the
  presence of the magnetic field $\bB=\nabla\times\bA$ and 
$\rho_{\br\sigma}= f_{\br\sigma}^\dagger f_ {\br\sigma}$ is the
electron number on site $\br$ with spin $\sigma$.

To specify the flake shape, we start with a circle divided into a
number $P$ of wedges, each of which has radius randomly chosen between
$R_-$ and $R_+$. This procedure generates a compact shape with an
irregular boundary.   The graphene tight-binding model is then implemented
on the resulting  shape using the Kwant Python package \cite{groth2014kwant}. 

Interactions are described by 
\begin{equation}\label{int2}
H_{\rm int} = {1\over 2} \sum_{\br,\br'}\rho_\br V(\br-\br')
                \rho_{\br'}, 
\end{equation}
where $\rho_\br=\rho_{\br\uparrow}+\rho_{\br\downarrow}$ represents
the total charge on site $\br$ and $V(\br)=(e^2/\epsilon r)e^{-r/\lambda_{TF}}$ is the
screened Coulomb potential.

Our strategy is to first solve the non-interacting problem defined by
$H_0$ on a flake with an irregular boundary. This yields a set of
single-particle energy levels $\epsilon_j$ and the corresponding
eigenstates $\Phi_j(\br)$. As already discussed in the main text the
energy levels consist of bulk Landau levels and edge modes. The
Zeeman term simply offsets the spin-up bands by $\Delta
E_S(B)=g^*\mu_BB$ with respect to spin-down bands. 

Next we write the
interaction term $H_{\rm int}$ in the basis defined by the eigenstates $\Phi_j(\br)$.
If $c_{j\sigma}^\dagger$ creates a particle with spin $\sigma$ in
eigenstate  $\Phi_j(\br)$ we have
$\rho_\br=\sum_{i,j,\sigma}\Phi_i^*(\br)\Phi_j(\br)
c_{i\sigma}^\dagger c_{j\sigma}$. Substituting into Eq.\ (\ref{int2})
and rearranging we find
\begin{equation}\label{int3}
H_{\rm int} =\sum_{i,j,k,l}\sum_{\sigma,\sigma'}J_{ij;kl}c_{i\sigma}^\dagger c_{j\sigma'}^\dagger c_{k\sigma'}c_{l\sigma},
\end{equation}
where $J_{ij;kl}$ is given by Eq.\ (3) in the main text.

Henceforth we focus on the states belonging to LL$_0$, that is, we
consider electron densities such that all Landau levels with negative
energies are filled, while LL$_0$ is partially filled. Given the LL
degeneracy $N=SB/\Phi_0$ per spin we define the total number of LL$_0$ electrons $N_F$
such that $N_F=0$ and $N_F=2N$ correspond to completely empty or
filled LL$_0$, respectively. Because higher LLs are separated by an
energy gap, for sufficiently weak interactions we can disregard
virtual transitions into these bands and project  $H_{\rm int}$ onto
LL$_0$ by simply restricting all indices $(i,j,k,l)$ in Eq.\
(\ref{int3}) to those labeling eigenstates $\Phi_j$ in LL$_0$.

\subsubsection{Exchange splitting}

We expect electrons to occupy LL$_0$ in such a way as to maximize the
total spin $\bS_{\rm tot}$ with $S^z_{\rm tot}$ aligned with the
field.  Such a state will minimize  the Zeeman energy
as well as the Coulomb repulsion due to the exchange effect. The
latter arises because when the spin part of the many-body electron
wavefunction is symmetric in spin degrees of freedom the spatial part
must necessarily be antisymmetric. This forces $\Psi(\br_1,\br_2,\dots)$ to vanish
whenever two electron positions coincide, which tends to minimize the
short-range part of the Coulomb repulsion energy. While the Zeeman
splitting is easy to determine (main text), estimation of the exchange
splitting magnitude for $N_F$ fermions described by Eq.\ (\ref{int3}) is a
non-trivial task. This is because couplings $J_{ij;kl}$ are all-to-all
and essentially random. To get an idea about the expected magnitude of
the exchange splitting we consider below a simple case of $N=N_F=2$.

For two electrons the position space wavefunction can be either symmetric or
antisymmetric under exchange depending on the spin state, 
$\Psi_\pm(\br_1,\br_2)={1\over
  \sqrt{2}}\left[\Phi_1(\br_1)\Phi_2(\br_2)\pm
  \Phi_1(\br_2)\Phi_2(\br_1)\right]$. The corresponding Coulomb energy is $E_C^\pm=\sum_{\br_1,\br_2}
|\Psi_\pm(\br_1,\br_2)|^2V(\br_1-\br_2)$. The exchange splitting,
then, becomes simply $\Delta E_C=E_C^+-E_C^-$ and reads
\begin{equation}\label{int4}
\Delta E_C=2\sum_{\br_1,\br_2}{\rm Re}\left[\Phi_1^*(\br_1)\Phi_2(\br_1)V(\br_1-\br_2)\Phi_2^*(\br_2)\Phi_1(\br_2)\right]
\end{equation}
In order to estimate $\Delta E_C$ from Eq.\ (\ref{int4}) we make an
assumption, motivated by our extensive numerical work,
that on lengthscales larger than the magnetic length $l_B=\sqrt{\hbar c/eB}$
wavefunctions $\Phi_j(\br)$ behave as random uncorrelated variables.
We thus coarse grain the wavefunctions on a grid with sites denoted by
$\bR$ and spacing $l_B$. The coarse-grained wavefunctions
$\Phi_j(\bR)$ are then treated as complex-valued  independent  random
variables with 
\begin{equation}\label{int5}
\overline{\Phi_j(\bR)}=0, \ \ \ \
\overline{\Phi_i^*(\bR)\Phi_j(\bR')}={1\over M_s}\delta_{ij}\delta_{\bR\bR'}.
\end{equation}
Overbar denotes
averaging over independent realizations of the flake geometry. The second
equality in Eq.\ (\ref{int5}) follows from the normalization of
$\Phi_j$ and
\begin{equation}\label{int55}
M_s=S/l_B^2=2\pi N
\end{equation}
denotes the number of grid sites in the flake.
\begin{figure*}
\includegraphics[width = 17.6cm]{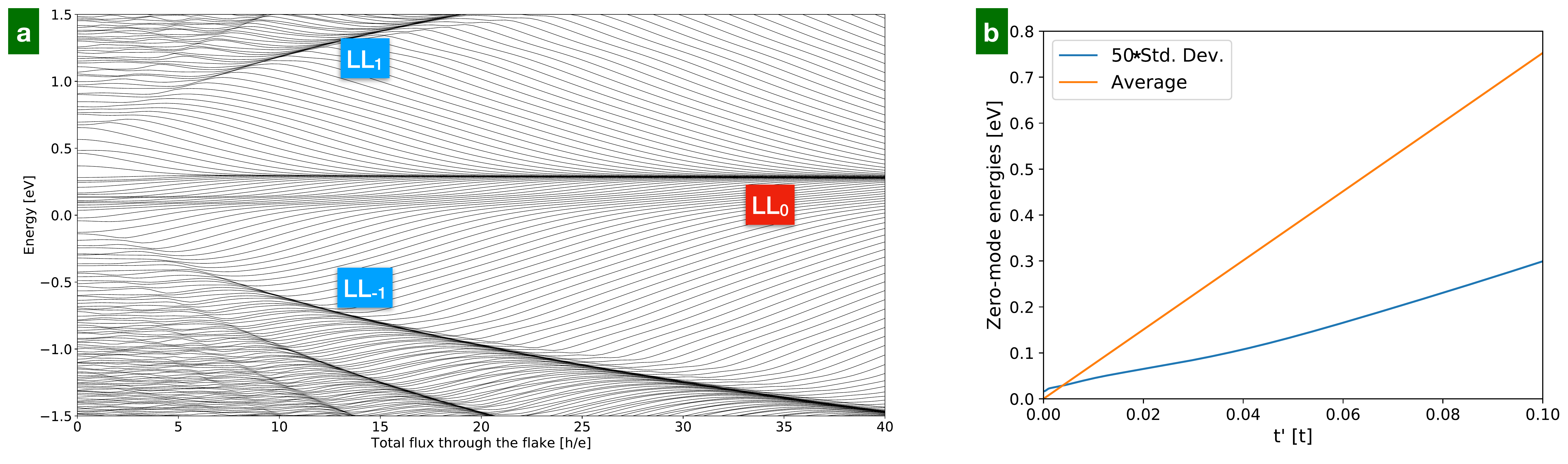}
\caption{{\bf Effects of the second neighbor hopping $t'$.} a) Single-particle
  energy spectrum of a flake (the same geometry as Fig.\ 2 main text)
  with second neighbor hopping $t'=0.037t$. b)  Average shift
  $\delta{\epsilon}=\overline{K_{ij}}$ 
and  standard deviation $K$ of 40 energy
levels that comprise LL$_0$ as a function of $t'$.
}\label{fig6}
\end{figure*}

With this preparation we now recast Eq.\  (\ref{int4}) as a sum over
the coarse grained grid, $\sum_{\br_1,\br_2}\to \sum_{\bR_1,\bR_2}$ and
$\Phi_j(\br)\to\Phi_j(\bR)$. Using Eq.\ (\ref{int5}) we then
obtain an estimate for the typical exchange splitting
\begin{equation}\label{int6}
\Delta
E_C\simeq{2\over M_s^2}\sum_{\bR_1,\bR_2}\delta_{\bR_1\bR_2}V(\bR_1-\bR_2)=
{2\over M_s}V(0).
\end{equation}
Here $V(0)$ must be interpreted as the average Coulomb potential in a grid patch of the
size $l_B$, that is $V(0)\simeq (1/\pi
l_B^2)\int_0^{l_B}V(r) 2\pi rdr=2e^2/\epsilon l_B$, where we assumed
$\lambda_{TF}\gg l_B$. Taking the dielectric
constant $\epsilon =2$ and $N=2$ we find the typical exchange
splitting $\Delta E_C\simeq 8.8$ meV/T. We expect this result to
remain at least approximately valid for $N>2$. Therefore, when
$N_F<N$, electrons will fill the spin-down states of LL$_0$ with empty
spin-up states separated in energy by a significant exchange
gap. The physics of such partially filled spin-down LL$_0$ can be
described by the Hamiltonian  (\ref{int3}) with $\sigma=\sigma'=\downarrow$
which is precisely the SY Hamiltonian.

\subsubsection{Coupling strength $J$}

To estimate the typical strength of couplings $J_{ij;kl}$ that enter
the SY Hamiltonian it is useful to first recast
Eq.\ (4) of the main text such that it is explicitly antisymmetric in
indices $(i,j)$ and $(k,l)$
\begin{equation}\label{int7}
J_{ij;kl}={1\over 2}\sum_{\bR_1,\bR_2}\Omega_{ij}^*(\bR_1,\bR_2)V(\bR_1-\bR_2)\Omega_{kl}(\bR_1,\bR_2),
\end{equation}
where
$\Omega_{ij}(\bR_1,\bR_2)={1\over 2}[\Phi_i(\bR_1),\Phi_j(\bR_2)]$. We
also passed to the coarse-grained variables, as described above. With help of Eq.\
(\ref{int5}) it is straightforward to show that
$\overline{J_{ij;kl}}=0$ and 
\begin{equation}\label{int8}
\overline{|J_{ij;kl}|^2}={1\over M_s^3}\sum_{\bR\neq 0}V(\bR)^2.
\end{equation}
The sum can be approximated by an integral,
\begin{equation}\label{int9}
\int_{l_B\over 2}^\infty{2\pi R dR\over l_B^2}\left({e^2\over 
    \epsilon}{e^{-R/\lambda_{TF}}\over R}\right)^2=
\left({e^2\over \epsilon l_B}\right)^2 2\pi \Gamma(0,{l_B\over \lambda_{TF}}),
\end{equation}
where $\Gamma(0,x)=\int_x^\infty dy \ e^{-y}/y$ is the incomplete gamma
function. Combining with Eq.\ (5) in the main text we thus obtain an
estimate 
\begin{equation}\label{int10}
J\simeq  2
\left({e^2\over\epsilon l_B}\right) \left(N\over M_s\right)^{3/2}\sqrt{\pi \Gamma(0,{l_B\over \lambda_{TF}})}.
\end{equation}
For $\epsilon=2$ this amounts to 
\begin{equation}\label{int11}
J\simeq 6.04 \  {\rm meV}
\sqrt{B[{\rm T}] \ \Gamma(0,{l_B\over \lambda_{TF}})}.
\end{equation}
For $x=l_B/\lambda_{TF}\ll 1$, which is the limit of interest,
$\Gamma(0,x)\simeq \ln(1/x)$ so $J$ is only very weakly dependent on
the screening length. For $B=20$ T and $\lambda_{TF}/ l_B=4$ we obtain
$J\simeq 25$ meV. 

It is to be noted that our numerical calculations of $J_{ij;kl}$
described in the main text [discussion below Eq.\ (5)] give larger
values of $J$ than the above estimate, in some cases by as much as an
order of magnitude. The discrepancy is most likely attributable to
the fact that LL$_0$ wavefunctions are in fact disordered on a somewhat
longer lengthscale than $l_B$. This would modify the relation between
$M_s$ and $N$ given by Eq.\ (\ref{int55}) and increase the ratio
$(N/M_s)$ that enters the estimate for $J$ in Eq.\ (\ref{int10}). We may
therefore regard  Eq.\ (\ref{int10}) as a conservative lower
bound on the expected magnitude of $J$.
This is already a large energy scale which should
make the manifestations of the SY physics experimentally observable at
low temperatures in clean graphene flakes.

\subsection{Symmetry breaking perturbations}

\begin{figure}
\includegraphics[width = 7.6cm]{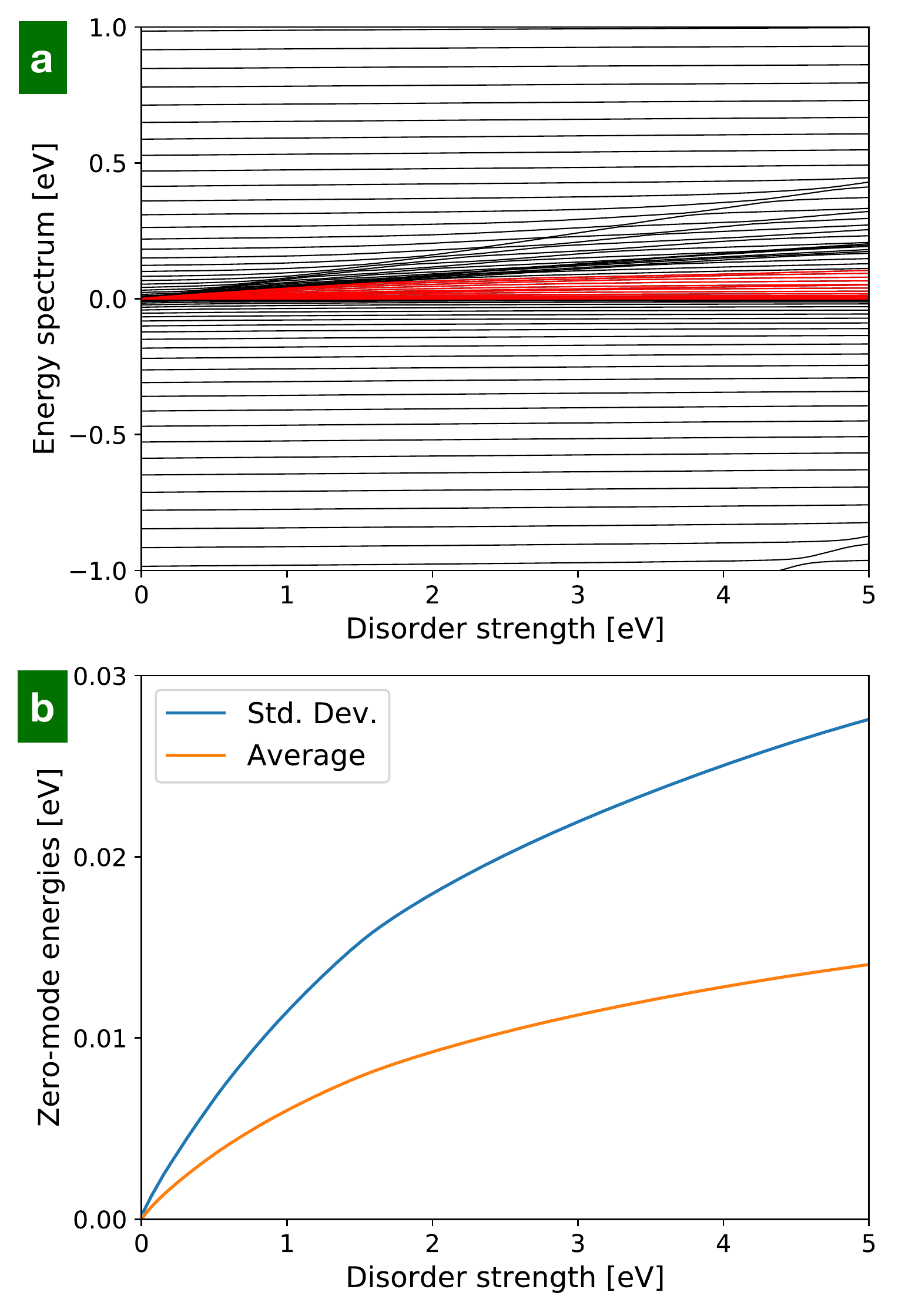}
\caption{{\bf Effects of random on-site potential.} a) Low-energy part
  of the numerically calculated energy spectrum for the flake  with
  $n_I=1$\% of defected sites as a function of the disorder potential
  strength $w$ and $N=40$. b)  Average shift $\delta{\epsilon}=\overline{K_{ij}}$ and 
standard deviation $K$ of 40 energy
levels that evolve from the zero modes which comprise LL$_0$ in the
pure sample. These levels are marked in red in panel (a).
}\label{fig5}
\end{figure}
To ascertain the experimental feasibility of our proposal we now
discuss the effect of various chiral symmetry breaking perturbations
that exist in real graphene. Such perturbations tend to broaden LL$_0$
and can be modeled by a bilinear term $\cH_2$ defined by Eq.\ (3) in
the main text. The matrix elements are 
\begin{equation}\label{dis0k}
K_{ij}=\sum_\br\Phi_i^*(\br)H'(\br)\Phi_j(\br), 
\end{equation}
where $H'$ denotes the Hamiltonian of the perturbation.
 The strength of these perturbations is measured by
parameter $K$ defined as  
\begin{equation}\label{dis0}
K^2=N\left(\overline{ |K_{ij}|^2}-|\overline{ K_{ij}}|^2\right). 
\end{equation}
It is known that since $\cH_2$ is a relevant perturbation  to 
$\cH_{\rm  SY}$  (in the
renormalization group sense) the ground state of the system becomes a (disordered) Fermi
liquid for any nonzero $K$. Nevertheless, if $K\ll J$, a significant
crossover region can exist at finite frequencies and temperatures in
which the system behaves effectively as a maximally chaotic SY liquid.
According to the analysis of Ref.\ \onlinecite{Pikulin2017} the
zero-temperature propagator of the system with both $K$ and $J$ nonzero
exhibits the SY conformal scaling $G(\omega)\sim |\omega|^{-1/2}$
for frequencies satisfying 
\begin{equation}\label{dis00}
16\sqrt{\pi}K^2/J <\omega\ll J.
\end{equation}
In the following
we consider two specific perturbations that are present in real
graphene, the second neighbor hopping $t'$ and random on-site
potential. Both break the chiral symmetry $\chi$ and produce non-zero
parameter $K$. We derive limits on the admissible strength of these
perturbations based on the requirement that Eq.\ (\ref{dis00}) yields
a significant window in which SY behavior can be observed.

\subsubsection{Second neighbor hopping}
We first consider second neighbor hopping with the Hamiltonian acting
as $H'(\br)\Phi_j(\br)=t'\sum_{\ba}\Phi_j(\br+\ba)$. Here $\ba$
denotes the 3  second neighbor vectors in the honeycomb lattice. Since
$|\ba|\ll l_B$ we find, upon coarse graining the sum in Eq.\
(\ref{dis0k}), 
\begin{equation}\label{dis1}
K_{ij}\simeq 3t'\sum_\bR\Phi_i^*(\bR)\Phi_j(\bR).
\end{equation}
With help of Eq.\ (\ref{int5}) it is straightforward to show that 
\begin{equation}\label{dis2}
\overline{K_{ij}}\simeq 3t'\delta_{ij}, \ \ \ \ 
\overline{|K_{ij}|^2}\simeq 9t'^2\left(\delta_{ij}+M_s^{-1}\right).
\end{equation}
From Eq.\ (\ref{dis0}) we get $K\simeq 3t'\sqrt{N/M_s} \simeq 3t'/\sqrt{2\pi}$ independent of
the field.

Experimentally reported values of $t'$ range between\cite{Neto_RMP} 1-3\% of $t$
which would produce a rather large broadening of LL$_0$ in real
graphene, $K\simeq 30-90$ meV. On the other hand existing experiments \cite{Song2010}
indicate much smaller broadening of Landau levels in graphene of at
most several meV 
which
also includes broadening due to impurities and other defects. We
therefore conclude that the above method must severely overestimate
the contribution of second neighbor hopping to parameter $K$. This
conclusion is supported by our numerical results presented below.    

The numerically computed energy spectrum of the graphene flake with second neighbor hopping $t'=0.037 t$ is
displayed in Fig.\ \ref{fig6}a. We observe that while LL$_0$ is now
significantly shifted away
from zero energy it remains sharp and well defined. The overall upward
shift of LL$_0$ by about 0.25 eV is consistent with the estimate given
in Eq.\ (\ref{dis2}) which implies $\overline{K_{ij}}\simeq 0.30$ eV.  The broadening
induced by $t'$ is quantified  in  Fig.\ \ref{fig6}b and is well
approximated by a linear dependence $K\simeq 0.022 t'$. This is about
a factor of 50 smaller than the estimate implied by Eq.\ (\ref{dis2}).
 For $t'=0.02t$ we obtain $K\simeq 1.2$ meV,  a result that is  much
 more in line with the experimental data. 

The discrepancy between the
 analytical estimate and the numerical result can be understood as
 follows. In a large, disorder-free sample of graphene, inclusion of the
second neighbor hopping produces changes in the band
 structure (and thus the position and spacing of LLs) but does not
 give rise to any LL broadening as long as $t'$ remains spatially
 uniform. The sharpness of LLs is protected by translational
 invariance, not the chiral symmetry. In our mesoscopic flake we see that
the  inclusion of a spatially uniform $t'$ primarily shifts the position
of LLs, as expected from the argument given above. Because
randomness is present in the system due to its irregular geometry
some broadening occurs. This broadening is, however, much weaker than what is
predicted by the naive estimate.

\subsubsection{Random on-site potential}
Random on-site potential is implemented by taking $H'(\br)=w\sum_{\br'\in {\cal I}}\delta_{\br\br'}$,
where $\cal I$ denotes a set of randomly chosen sites with number density $n_I$ in the graphene
lattice and $w$ controls the disorder potential strength. Substituting
$H'$ into Eq.\ (\ref{dis0k}) leads to the same result as indicated in
Eq.\ (\ref{dis2}) with $3t'$ replaced by $wn_I$. We therefore expect
an overall energy shift of LL$_0$ by $wn_I$ accompanied by a
broadening
\begin{equation}\label{dis3}
K\simeq wn_I\sqrt{N\over M_s}.
\end{equation}

Fig.\ \ref{fig5}a shows the numerically computed 
energy eigenvalues as a function of $w$ for a flake with
$N=40$ flux quanta and $n_I=1$\%. We observe that LL$_0$ is shifted upward
as well as broadened with increasing disorder strength. This shift
$\delta\epsilon$ and  broadening $K$ are
quantified in Fig.\ \ref{fig5}b. At small $w$ these satisfy
$\delta\epsilon\simeq 0.8 n_Iw$ and
$K\simeq 1.3 n_Iw$ while at larger values the dependence is no longer
linear, presumably because the system enters a non-perturbative
regime when $w$ becomes comparable to the bandwidth. We see that the
numerically obtained shift in LL$_0$ is well aligned with the
analytical estimate. The broadening $K$ also agrees if we take
$\sqrt{N/M_s}\simeq 1.3$ (instead of $1/\sqrt{2\pi}\simeq0.4$ implied by Eq.\  \ref{int55}).
This result reinforces the conclusion,  reached in Appendix A by 
comparing the interaction strength estimate to the numerical
calculation,  that the zero mode wavefunction
disorder scale is somewhat longer than $l_B$.

We finally remark that in the above example 40 flux quanta through a flake with
1952 carbon atoms correspond to an unrealistically high magnetic
field of $\sim 3200$ T. Such high fields are needed for us to be able
to numerically simulate meaningful number of zero modes $N$ with
available computational resources. To make a closer contact with
experiment we may however reinterpret these results by viewing the honeycomb
lattice not as the atomic carbon lattice but as a convenient regularization of the low energy theory of
Dirac electrons in graphene. In such low energy theory the only
important parameter is the Dirac velocity $v_F={3\over 2}ta\simeq
1.1\times 10^6$ m/s. The velocity is clearly unchanged if we rescale the
lattice constant $a\to \lambda a$ and the tunneling amplitude $t\to
t/\lambda$ with $\lambda$ an arbitrary positive
parameter. Under the rescaling $B\to B/\lambda^2$ and all energy
parameters defined through $t$ are changed as $E\to E/\lambda$. Thus,
if we take $\lambda=10$ in the above example we get a more reasonable field
$B=32$ T. According to Eq.\ (\ref{int11}) this corresponds to $J\simeq 34$
meV.  Eqs.\ (\ref{dis3}) and (\ref{dis00}) then stipulate an upper
bound on the disorder strength $n_Iw\ll 9$ meV.  

Clearly, like
fractional quantum Hall effect and other exotic phases driven by
interactions, observing the SYK physics will require high fields, low
temperatures and carefully prepared graphene flake with an irregular
boundary and clean interior.

\end{document}